# Stochastic Cooling Overview

John Marriner
Fermi National Accelerator Laboratory[*]

Abstract

The status of stochastic cooling and developments over the years are reviewed with reference to much of the original work.  Both theoretical and technological subjects are considered.

## Development of the Concept

A comprehensive description of stochastic cooling has been given in the classic paper of Mohl, Petrucci, Thorndahl, and van der Meer.[1]  This paper has served as the basic reference for stochastic cooling, and contains all of the important results, except for a detailed theory of bunched beam cooling.  A fundamentally different theoretical approach that yields essentially the same results has been given by Bisognano.[2]  Other early work includes Derbenev and Kheifets,[3] Hereward,[4] Katayama,[5] Palmer,[6] and Sacherer.[7]  A number of authors have summarized the field from various points of view including Caspers,[8,9] Mohl,[10] and Marriner.[11]

## Stochastic Cooling Concepts

The general idea of stochastic cooling is to sample a particle's motion with a pickup and to correct the motion with a kicker.  Stochastic cooling is similar to other beam feedback systems used on accelerators, except that the stochastic cooling system works on individual (incoherent) particle amplitudes, not the (coherent) motion of the beam as a whole.  No stochastic cooling system is able to resolve individual particles in a single sample, but after a sufficiently long time (the cooling time) each particle develops its own dissipation (damping) force in a sea of much larger signals from the sum of all the other particles.  The key point is that every particle has a slightly different frequency of motion, and the force generated by all the other particles has a random phase and thus averages to zero.  The net result is that cooling of each particle can be described by a damping force, which is created by the particle and is linear in the system feedback gain, and the heating force, which is created by all the other particles and averages to zero to first order in the feedback gain but causes particle diffusion proportional to the gain squared.  A more detailed, pedagogical explanation of the principles can be found in Marriner and McGinnis[12] or Mohl.[13]

[*]Work supported by the US Department of Energy (DOE) under contract number DE-AC02-76CH03000.

*Transverse Cooling*

Transverse cooling is achieved by sensing the particle displacements in the pickup and applying a correcting signal at the kicker. Normally, the pickup and kicker are placed ninety degrees apart in betatron phase so that a position displacement at the pickup will become an angular displacement at the kicker. A simplified schematic of the process is shown in Figure 1.

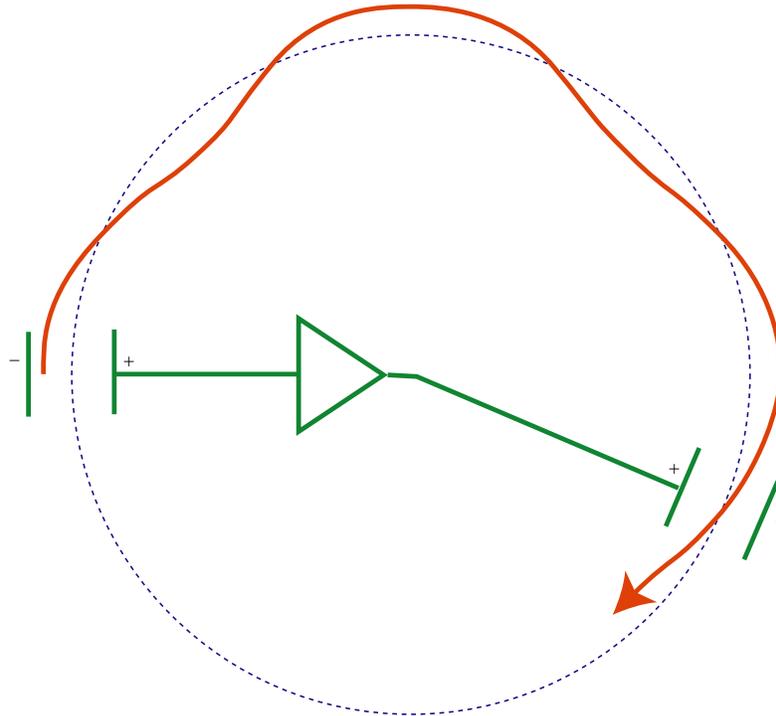

Figure 1. A cartoonist's view of a transverse stochastic cooling system.

Filters can be, and have been, used in transverse systems to cool beams. The filter stores the beam signal and corrections are made in subsequent turns with the correct phase. The phase response of the filter can be used to compensate for a pickup to kicker phase advance that differs from 90º. One implementation used for the FNAL Debuncher had notches at multiples of half the revolution frequency and eliminated the revolution harmonic signal (from incomplete common mode rejection) and half the thermal noise. In this case, the betatron sideband phase was unaffected because the filter phase shift was zero at the Schottky sidebands (the fractional tune was near ¼).

*Momentum Cooling*

There are two basic types of momentum cooling. One technique was suggested by Robert Palmer and is often referred to as Palmer cooling. This technique relies on the correlation between position and momentum in regions of high dispersion. A difference pickup makes a zero in the gain function at some momentum where the beam accumulates. The kicker is usually placed in a zero dispersion region, and equally kicks beam particles of all momenta. Since beam particles exhibit a long-term response only to excitations at their individual revolution frequencies, there is no disadvantage to having the kicker in zero dispersion.

Another technique, known as filter cooling,[14] relies on the correlation between revolution frequency and momentum. In this technique a pickup that is equally sensitive to all beam particles is used, but a zero in the electronic gain is necessary at each Schottky band. The zeros are produced with a notch filter that has a period precisely equal to the target revolution frequency. The system is built so that the real part of the gain changes sign around the notch frequency, as is required for cooling.

Palmer cooling is advantageous because it avoids the extra phase shifts of notch filters, an important advantage when the Schottky bands are wide. Filter cooling is advantageous in low noise situations because it notches the thermal noise as well as the Schottky signal. Palmer systems are generally more stable, a critical factor for high dynamic range stacking systems.

Another technique, which is sometimes called transit time cooling, is a special case of filter cooling. It is based on using the "bad mixing" or transit time effect to change the sign of the real part of the gain function around the particle revolution frequency. Under normal conditions, the transit time effect changes the sign of the imaginary part of the gain. However, if a 90° phase shift is inserted—by differentiation of the gain function, for example—a cooling effect may be obtained. Transit time cooling, however, does not provide any advantage compared to Palmer or filter cooling except possibly for easier implementation.

## Stochastic Cooling Implementations

The first practical cooling system was demonstrated at the ISR.[15,16] Early R&D work was done in storage rings dedicated to that purpose: ICE[17] at CERN and the 200 MeV electron cooling ring (ECR) at FNAL.[18,19] The R&D proved the stochastic cooling technique – particularly the critical demonstration of simultaneous stochastic beam cooling in all three planes – and enabled the confident construction of the CERN AA[20,21] and later the two-ring Fermilab antiproton source including the Debuncher[22] and the Accumulator[23] Rings. The ACOL ring was later added to the CERN antiproton source[24] and the Fermilab Debuncher cooling system bandwidth was increased.[25] Other early experimental cooling systems included TARN[26] and NAP-M.[27]

Low energy antiproton systems were developed for the ISR[28] and the dedicated storage ring LEAR.[29,30,31] The ISR antiproton program was continued at the Fermilab Accumulator. The Antiproton Decelerator[32] (AD) was later built at CERN as a low-cost successor to LEAR. More recently there has been an interest in applying stochastic cooling techniques to the proton and heavy ion beams in the medium energy region at COSY[33,34] and ESR.[35,36]

A summary of the major stochastic cooling systems is shown in Table 1. In general, one sees applications to improve source brightness and accumulate particles and also to improve the rate of beam interaction with a target. Many of the systems have been modified several times in a way that is difficult to capture in a simple table. The references should be consulted for the details.

## Bunched Beam Cooling

The theory of bunched beam cooling was developed in a comprehensive way by Bisognano and Chattopadhyay.[37,38] Bunched beam cooling has been successfully demonstrated in the Fermilab Accumulator.[39] More Recently, cooling of a beam captured in a barrier bucket has been demonstrated at the Recycler Ring at Fermilab.[40] However, attempts to extend the technique to high-energy colliding beams at the CERN $Sp\bar{p}S$[41] and the Fermilab Tevatron[42] have not proven successful. A summary of the problems associated with bunched beam cooling has been given by Caspers and Mohl.[43] As these authors noted, a major problem is that bunched beams tend to show unexpectedly large coherence at microwave frequencies. This coherence appears to be a collective self-bunching of the beam. A serious effort is underway to develop the technique at RHIC[44,45] where coherent effects seem to be less severe.

Table 1. A list of stochastic cooling systems and basic parameters.

| Site | Machine | Type | Frequency (MHz) | Beam Momentum (GeV/c) |
|---|---|---|---|---|
| CERN | ISR | H & V | 1000-2000 | 26.6 |
| | ICE | H, V, ΔP | 50-375 | 1.7 & 2.1 |
| | AA | PreCool ΔP<br>ST H, V, ΔP<br>Core H, V, ΔP | 150-2000 | 3.5 |
| | LEAR | 2 systems<br>H, V, ΔP | 5-1000 | <0.2 &<br>0.2-2.0 |
| | AC | H, V, ΔP | 1000-3000 | 3.5 |
| | AD | H, V, ΔP | 900-1650 | 2.0 & 3.5 |
| FNAL | ECR | V, ΔP | 20-400 | 0.2 |
| | Debuncher | H, V, ΔP | 4000-8000 | 8.9 |
| | Accumulator | ST ΔP<br>Core H, V, ΔP | 1000-8000 | 8.9 |
| KFA Julich | COSY | H, V, ΔP | 1000-3000 | 1.5-3.4 |
| GSI Darmstadt | ESR | H, V, ΔP | 900-1700 | 0.48/nucleon |
| Tokyo | TARN | ΔP | 20-100 | 0.007 |
| BINP | NAP-M | ΔP | 100-300 | 0.062 |

**Bandwidth**

The rate of cooling depends on the cooling system bandwidth. The bandwidth is limited by its highest frequency: the highest frequency utilized to date is 8 GHz. Aside from technological issues, a complicating factor is that the beam pipe aperture tends to be comparable to the wavelength at high frequencies, implying that the pickup and kicker structures as well as the beam pipe itself support many electromagnetic modes. It appears to be fairly easy to absorb energy traveling in the beam pipe: an early example was reported by Barry.[46] The pickup and kicker design problem has not been fully solved although there has been a lot of work on differing approaches. The most promising application for very high frequency systems, namely high-energy accelerators where the beam size is small, has been stalled because these applications involve the problematic cooling of bunched beams.

There has been significant interest in developing stochastic cooling at optical frequencies to obtain a bandwidth of perhaps 10,000 GHz.[47] However, a practical system has not yet been demonstrated.

## Power Sources & Limitations

Stochastic cooling systems require broadband power amplifiers at power levels ranging from Watts to 1000's of Watts. Solid state amplifiers have been used routinely for frequencies below 1 GHz. The CERN ACOL project developed high power solid state amplifiers[48] specifically for the 3 stochastic cooling bands in the 1 to 3 GHz frequency range. Traveling wave tubes[49] (TWT's) have been used for high power applications in the 1 to 8 GHz band, most extensively at FNAL.

The performance of some stochastic cooling systems is limited by the total power available, not by the system bandwidth. This situation has been analyzed by Goldberg, et al.,[50] who suggest that higher pickup sensitivity can be more important in these cases than an increase in bandwidth.

## Pickup Technologies

A wide variety of pickup designs have been used in stochastic cooling systems. A general description of pickup and kicker principles has been given by Goldberg and Lambertson.[51] Realistic designs have generally been optimized for varying beam energies, beam apertures, and other local considerations. However, pickups can be divided into two general categories.

*Phased arrays*

Phased arrays consist of many individual pickups of relatively low impedance. The outputs of a series (up to 100's) of pickups are added to achieve the desired sensitivity. The mostly widely used structure is a stripline (also known as a "loop") pickup. These pickups are generally constructed from a plate placed parallel to the beam and the vacuum chamber with connections at the upstream and downstream end of the plate. Originally made from sheet metal,[52,53] they can also be made on printed circuit boards,[54,55,56] which offer the advantage of more economical construction. One advantage of using a large array of pickups is that it is relatively easy to adapt the array to a varying beam velocity.

*Traveling wave structures*

In a traveling wave, the beam induces a wave that travels at the beam velocity and grows as the wave and the beam travel the length of the structure. Maintaining synchronism between the beam and the traveling wave results in a tradeoff between bandwidth and sensitivity in these devices. The signals from traveling wave pickups can be added, but often a single unit has enough sensitivity. Relatively tight constraints on synchronism make it difficult to adapt this type of pickup to varying beam velocities.

Faltin[57] developed a slotted transmission line that was used at the ISR and later at the CERN AA; McGinnis[58] developed a somewhat similar device but based on slotted waveguide. Lower velocity pickups using a helical structure were developed for TARN[59] and the FNAL electron cooling ring, where a traveling wave transverse pickup was also used.

*Sensitivity calculations*

Calculating the response of pickup and kicker structures from first principles has been a daunting task for all but the simplest of structures. However, recent work has made significant progress towards reliable numerical design calculations for practical pickup arrays.[60,61,62] It should be noted that the traveling wave structures (references 57 and 58) are accurately described by semi-analytic models that has been verified by direct measurement.

*Cryogenics*

Stochastic cooling pickups are typically back-terminated with the characteristic system impedance in order to achieve a high bandwidth. The terminating resistors are a source of noise proportional to the resistor temperature (Johnson noise). This noise can be dramatically reduced by cooling the resistors to cryogenic temperatures. The amplifier noise is also generally reduced by cooling.[63] Another benefit of cryogenically cooling pickups and amplifiers is that the resistance of copper conductors is reduced by cooling. Signal loss (though generally not large even at room temperature), can be reduced significantly in cryogenically cooled systems.

*Other pickup sensitivity considerations*

The ability to combine and split microwave signals without loss of power is an important consideration in stochastic cooling systems. Large phased arrays of pickups, especially, require an efficient power combiner.[64] Since the signal to noise ratio is critical in many systems, some pickups have been built with an adjustable gap that decreases in size as the beam cools.[65]

## Filter Technology

There have been a number of technical realizations of notch filters suitable for use in stochastic cooling. The realization of stochastic cooling filters is demanding since the uniformity of spacing must be high and the notches must be deep. A key element of all notch filters is an element that has a precise, frequency independent delay and a low attenuation. A variety of circuits are possible, varying in technical complexity and response.[66] Many filters have used low loss transmission lines as the basic element. More recent variants include the use of novel delay elements in analog circuits: superconducting transmission lines,[67] bulk acoustic wave devices,[68] and optical fibers.[69,70] The optical fiber is implemented by modulating the intensity of a laser with the microwave signal, transmitting the signal over an optical fiber, and demodulating the light received. In addition to analog techniques, advances in digital signal processing have made it relatively easy to realize arbitrarily complicated but very precise filters,[71] particularly for the lower frequencies used in stochastic cooling systems.

## Gain

The gain of a stochastic cooling system is characterized by the real and imaginary parts of the beam transfer function (open loop gain) as a function of frequency. At the design stage the gain is usually modeled as constant over a given frequency band and zero outside this band, although such a gain function cannot be realized in practice. More realistic gain functions have been studied theoretically by van der Meer.[72] In practice, however, cooling systems are built with a target bandwidth and, after the system response

has been measured, modified by the addition of an equalizer to maximize the cooling rate.[73]

## System Measurements

The most important tool for understanding system performance is the open loop gain or beam transfer function (BTF) measurement. The BTF measurement yields a full characterization of a linear system. The measurement is made by breaking the connection between the pickup and kicker at any convenient point. A network analyzer is used to excite the kicker side of the broken connection, while the response is measured on the pickup side. The response of the beam can be accurately predicted by the knowledge of a few beam and lattice parameters (beam energy and spread, transition energy, and tune) and the response of the system (complex gain as a function of frequency) can therefore be readily determined. In practice, it is unnecessary to determine the beam response: it is enough to know that the peak responses in different Schottky bands have the same phase and an amplitude response that is inversely proportional to frequency. Generally, the desired response is one that is flat in both amplitude and phase.

An example of a BTF measurement from the FNAL Accumulator transverse core cooling system is shown in Figure 2abcd. Figure 2ab shows the amplitude and phase response at a single harmonic ($h=4305$). The plots show a characteristic peaking at the two betatron sidebands ($n\pm Q$) and a rapid change of phase of 180º in the vicinity of the resonance. The data (dots) is compared to the theoretical calculations of the beam response assuming that the cooling system gain is constant over this Schottky band. The predictions shown are a numerical integration of the Schottky spectrum measured using a resonant 79 MHz pickup and a gaussian fit to the same data. Except for the noisy nature of the numerical integration, the curves are virtually indistinguishable. The wide band response is shown in Figure 2cd. To make this measurement, the peaks on the betatron sidebands are sampled every few Schottky bands, and the response is interpreted in terms of the desirability of the stochastic cooling electronic gain function. One notices that the amplitude response is fairly flat over the range of 1600 to 4000 MHz, quite good for this nominally 2-4 GHz cooling system. Figure 2d plots the difference between the measured phase and the desired phase of 180º. The phase is likewise quite flat over this region, but the phase is slightly too high. Lengthening the system delay by about 10 psec will fix this problem. The difference between upper and lower sideband phase is equal to twice the error in phase advance. This error is zero for the measurement shown in Figure 2d.

## Conclusion

Stochastic cooling is a mature subject that has been successfully applied to the problem of creating intense particle sources (most spectacularly for antiprotons) and for reducing beam size, especially in the energy regime where electron cooling has traditionally been judged to be impractical or at least more difficult. One outstanding question is whether increases beyond the maximum achieved frequency of 8 GHz are possible and, in particular, whether cooling at optical frequencies will be practical. In addition, the application of stochastic cooling to bunched beams in a colliding beam accelerator has proved so far to be elusive. Work on practical techniques has been steady especially in the area of pickup and filter technology. Advances in microwave technology and the application of digital techniques to rf systems have also influenced system design.

New applications of stochastic cooling for ions and modest energy proton beams have more recently appeared at COSY and ESR. It will be interesting to see how these and other possible future applications utilize stochastic cooling given potentially competing techniques such as higher energy electron cooling.

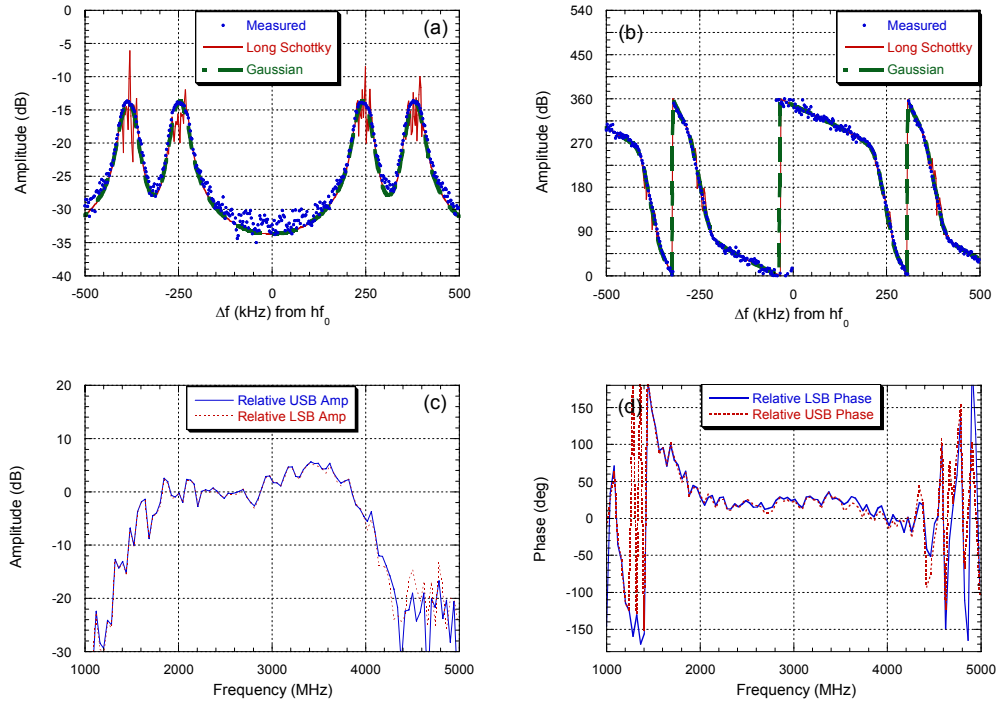

Figure 2. A network analyzer open loop gain (BTF) measurement of the Fermilab Accumulator 2-4 GHz stochastic cooling system. The amplitude (a) and phase (b) response of a single Schottky band at harmonic 4305 is shown (dots). The curves are calculations of the response using a measurement of the longitudinal Schottky beam profile. The amplitude (c) and phase (d) response sampled over the system bandwidth are also shown.

# References

[1] D. Mohl, G. Petrucci, L. Thorndahl, and S. van der Meer, Phys. Rept. **58** (1980), 76.
[2] J. Bisognano, IEEE Trans. Nucl. Sci. **30** (1983), 2393.
[3] Ya S. Derbenev and S.A. Kheifhets, Part. Accel. **9** (1979), 237.
[4] H.G. Hereward, CERN/77-13 (unpublished).
[5] T. Katayama and N. Tokuda, Part. Accel. **21** (1987), 99.
[6] R.B. Palmer, BNL/CRISP/73-19 (unpublished).
[7] F. Sacherer, CERN/ISR/TH/78-11 and Proceedings of the Workshop on Producing High Luminosity High Energy Proton-Antiproton Collisions (Berkeley, 1978), 93.
[8] F. Caspers, Workshop on Crystallization and Related Topics (Erice, 1995), 111.
[9] F. Caspers, Workshop on Beam Cooling and Related Topics (Bad Honef, 2001) on CD-ROM and CERN/PS/2001/017/RF.
[10] D. Mohl, Nucl. Instrum. & Meth. **A391** (1997), 164.
[11] J. Marriner, Proceedings, Workshop on Beam Cooling and Related Topics (Montreux, 1993), 14.

[12] J.P. Marriner & D. McGinnis, AIP Conf. Proc. **249**, Eds. M. Month and M. Dienes, American Institute of Physics (New York, 1992), 693.
[13] D. Mohl, Procedings, Antiprotons for Colliding Beam Facilities (Geneva, 1983), 163.
[14] G. Carron, et al., CERN/ISR/RF/78-12 (unpublished).
[15] G. Carron, et al., IEEE Trans. Nucl. Sci. **25** (1977), 1402.
[16] P. Bramham, et al., Nucl. Instrum. & Meth. **125** (1975), 201
[17] G. Carron, et al., IEEE Trans. Nucl. Sci. **26** (1979), 3456.
[18] R.L. Hogrefe, et al, IEEE Trans. Nucl. Sci. **28** (1981), 2455.
[19] G.R. Lambertson, et al., IEEE Trans. Nucl. Sci. **28** (1981), 2471.
[20] S. van der Meer, IEEE Trans. Nucl. Sci. **28** (1981), 1994.
[21] S. van der Meer, Rev. Mod. Phys. **57** (1985), 689.
[22] D. Peterson, et al., Proceedings of PAC87 (Washington), 1728.
[23] R. Pasquinelli, et al., Proceedings of PAC87 (Washington), 1132 and W. Kells, et al., Proceedings of PAC87 (Washington), 1090.
[24] B. Autin, et al., Proceedings of PAC87 (Washington), 1549.
[25] D. McGinnis, Proceedings of PAC99 (New York), 59.
[26] N. Tokuda, et al., Proceedings of the $12^{th}$ Int. Conf. on High-Energy Accelerators (Batavia, 1983), 386.
[27] E.N. Dement'ev, et al., Sov. Phys. Tech. Phys. **27** (1982), 1225.
[28] E. Peschardt and M. Studer, IEEE Trans.Nucl.Sci. **30** (1983), 2584.
[29] S. Baird, et al., Proc. Physics at Lear with Low Energy Antiprotons (Villars-sur-Ollon, 1987), 175.
[30] F. Caspers, et al., Proceedings, Workshop on Beam Cooling and Related Topics (Montreux, 1993), 207.
[31] J. Bosser, Proceedings, Low Energy Antiproton Physics (Stockholm 1990), 486.
[32] C. Carli and F. Caspers, Proceedings of EPAC 2000, 2414.
[33] R. Stassen,, et al., Proceedings of EPAC 1998 (Stockholm), 553.
[34] D. Prasuhn, et al., Nucl. Instrum. & Meth. **A441** (2000), 167.
[35] F. Nolden, et al., Proceedings of EPAC 2000 (Vienna), 1262.
[36] F. Nolden, et al. Nucl. Instrum. & Meth. **A441** (2000), 219.
[37] J. Bisognano and S. Chattopadhyay, IEEE Trans.Nucl.Sci. **28** (1981), 2462.
[38] S. Chattopadhyay, IEEE Trans. Nucl. Sci. 30 (1983), 2652.
[39] J. Marriner, et al., Proceedings of EPAC 1990 (Nice), 1577.
[40] D. Broemmelsiek, in preparation for PAC03 (Portland).
[41] D. Boussard, Frontiers of Particle Beams (South Padre Island, 1986), 269
[42] G. Jackson et al., Proceedings of PAC93 (Washington), 2148.
[43] F. Caspers and D. Mohl, 17th Int. Conf. on High Energy Accelerators (Dubna, 1998), 397.
[44] J. Wei, Proceedings, Workshop on Beam Cooling and Related Topics (Montreux, 1993), 132.
[45] J.M. Brennan, Proceedings of EPAC 2002 (Paris), 308.
[46] W. Barry, IEEE Trans. Nucl. Sci. **32** (1984), 2424.
[47] A.A. Mikhailichenko and M.S. Zolotorev, Phys. Rev. Lett. **71** (1993), 4146.
[48] G. Carron, et al., CERN/PS/85-01 (unpublished).
[49] B. Leskovar and C.C. Lo, IEEE Trans. Nucl. Sci. **30** (1983), 3423.
[50] D.A. Goldberg, et al., Part. Accel. **30** (1990), 7.
[51] D.A. Goldberg and G.R. Lambertson, AIP Conf. Proc. **249**, Eds. M. Month and M. Dienes, American Institute of Physics (New York, 1992), 537.
[52] D.A. Goldberg et al., IEEE Trans. Nucl. Sci. **32** (1985), 2168.
[53] D. McGinnis, et al. Proceedings of PAC89 (Chicago), 639.
[54] F. Caspers, Proceedings of PAC87 (Washington), 1866.
[55] J. Petter, et al., Proceedings of PAC89 (Chicago), 636.
[56] D. McGinnis, et al., Proceedings of PAC91 (San Francisco), 1389.
[57] L. Faltin, Nucl. Instrum. & Meth. **148** (1978), 449.
[58] D. McGinnis, Proceedings of PAC99 (New York), 1713.
[59] N. Tokuda, Physics at LEAR with Low Energy Antiprotons (Villars-sur-Ollon, 1987), 135.
[60] R. Schultheis, Proceedings of PAC95 (Dallas), 2342.
[61] M. Balk, et al., Proceedings of EPAC 2002 (Paris), 2670.
[62] P. Raabe, et al., Proceedings of EPAC 1992 (Berlin), 919.
[63] B. Leskovar and C.C. Lo, IEEE Trans. Nucl. Sci. **30**, 2259.

[64] Jimmie K. Johnson, IEEE Trans. Nucl. Sci. **31**, 2171.
[65] P. Lebrun, et al., Proc. 1985 Cryogenic Engineering Conf. (Cambridge), 543.
[66] S. L. Kramer, IEEE Trans. Nucl. Sci. **30** (1983), 3651. See also reference 1.
[67] M. Kuchnir, et al., IEEE Trans. Nucl. Sci.**30** (1983), 3360.
[68] R. Pasquinelli, Proceedings of PAC91 (San Francisco), 1395.
[69] R. Pasquinelli, et al., PAC89 (Chicago), 694.
[70] U. Bechstedt, et al., Proceedings of PAC99 (New York), 1701.
[71] M. Chanel, et al., Proceedings of EPAC 1990 (Nice), 1574.
[72] S. van der Meer, CERN/PS/AA/83-48 (unpublished).
[73] D. McGinnis, Proceedings of PAC91 (San Francisco), 1392.